\documentstyle[aaspptwo]{article}

\begin{document}

\title{AN IMPROVED MEASURE OF QUASAR ORIENTATION}

\centerline{arch-ive/9505127}
\author{Beverley J. Wills and M. S. Brotherton}
\affil{McDonald Observatory and Astronomy Department, University of Texas,
 Austin, TX 78712\\Internet: bev@astro.as.utexas.edu, msb@astro.as.utexas.edu}

\centerline{Submitted 1995 April 16, to appear mid-July, ApJ Letters}
\begin{abstract}
Radio core dominance, usually measured by R, the rest frame ratio of
core to lobe flux density, has been used as an indicator of Doppler boosting
of quasars' radio jets and hence the inclination of the central engine's
spin axis to the line of sight.
The use of lobe flux density as a means of normalizing the boosted core flux
density to the available intrinsic power of the central engine introduces
scatter.  This is because the emission from the radio lobes depends strongly on
the interaction of the jet with the environment at distances beyond several
Kpc from the nucleus.  Total kinetic power in the extended radio emission is
approximately proportional to emission line luminosity, and emission line
luminosity is proportional to the luminosity of AGNs' featureless continua --
both over 4 orders of magnitude.  Thus, quasars' optical luminosity may be
an excellent measure of available intrinsic jet power.  Therefore we
define a new core dominance parameter, R$_V$, the ratio of radio core
to optical (rest frame V band) continuum luminosity, that is not directly
dependent on jet interactions with the super-Kpc-scale environment.  We show
that the use of R$_V$, rather than R, results in significantly improved inverse
correlations with the beaming angle as deduced from apparent superluminal
velocities and inverse Compton scattered X-ray emission, and with the FWHM of
quasars' broad H$\beta$ emission line.  We discuss some implications and
applications of the new parameter.

\end{abstract}
\vfil\eject
\section{Introduction}

Different viewing angles successfully unify a wide variety of sizes and
structures of extragalactic radio sources, their optical classification,
and many other properties.  Bright radio bright cores are believed to be
simply end-on views of extended double-lobed sources with electron synchrotron
emission from their weak nuclear jet Doppler-boosted $\sim 1000$-fold into a
narrow cone
making a small angle to the line of sight (e.g., Blandford \& Rees 1978; Orr \&
Browne 1982; Hough \& Readhead 1989).  In this model, the observed jet
direction
near the nucleus defines the projected rotation axis of a massive central
engine.  The angle of the axis to the line of sight may be measured by the
ratio R of flux density in a nuclear core, unresolved on arcsecond scales,
to that emitted essentially isotropically from the extended radio lobes.

Because the line- and continuum-emitting central regions of distant
quasars cannot be spatially resolved, the correct interpretation of radio
core dominance is important, as it provides one of the few available tools for
investigating the geometry of AGNs.  This orientation indicator provided the
basis for several statistical investigations showing axial beaming of optical
and X-ray continuum emission, kinematic axisymmetry of the broad-emission line
gas, probable axisymmetry of optical depth in emission line regions,
obscuration, and dust reddening (e.g., Browne \& Murphy 1987; Impey et al.
1991;
Wills et al. 1992; Wills \& Browne 1986; Jackson et al. 1989; Jackson \& Browne
1991; Baker et al. 1994; Brotherton 1995).

High resolution milliarcsecond maps of jet structure and superluminal motion
show relativistic velocities for knots in nuclear jets, with a typical Lorentz
factor $\Gamma \sim$ 5 (Vermeulen \& Cohen 1994).
Ghisellini et al. (1993) compared these apparent
expansion speeds for about 40 AGNs, with the Doppler factor derived from
inverse Compton scattering of radio photons into the X-ray regime.
They showed that a simple ballistic model fits the data, and that,
statistically, the Lorentz factor giving rise to these observed pattern speeds
is, within the uncertainties, the same as that for the bulk velocities of the
synchrotron-emitting electrons. (But see the detailed discussion by
Vermeulen \& Cohen 1994).  Then, assuming these two speeds to be equal,
they derived, for individual sources, $\Gamma$\ and $\phi$, the angle of the
beam to the line of sight.  Ghisellini et al. showed that R appears to be an
indicator of $\phi$.  An updated version of this result is shown in Fig. 1a.

Ideally, one would prefer to normalize the beamed core luminosity to the
intrinsic jet power.  The extended radio luminosity is not necessarily the best
measure of this.  Recently Bridle et al. (1994) presented high dynamic
range VLA maps for part of their sample of 3CR quasars, and concluded that,
while the one-sided inner few kiloparsecs are dominated by relativistic
beaming, the regions beyond are increasingly affected by interactions with a
clumpy intergalactic medium.  The lobe flux densities are as much an indication
of beam power losses via interactions with the interstellar or intergalactic
medium as they are an indication of the intrinsic power of the central engine.

\section{An Improved Core Dominance Parameter}

We propose a different way to normalize core luminosity using the optical
luminosity
as a measure of intrinsic jet power.  This is strongly justified by two results
in combination.
First, Yee \& Oke (1978) and Shuder (1981) showed that emission-line luminosity
is proportional to the luminosity of the featureless continuum over four orders
of
magnitude, for a range of AGN types, and including the broad emission lines of
radio galaxies and QSOs.  Second,
Rawlings \& Saunders (1991) found that emission-line luminosity is
approximately
proportional to the total jet kinetic power for an unbiased sample of FR II
radio galaxies, and that this relation holds over four orders of magnitude when
high
and low luminosity radio galaxies, and even broad lined quasars (z $<$\ 1), are
included.  They concluded that the luminosity of the ionizing radiation is
proportional to jet kinetic power, and that both are closely coupled to the
power of the central engine.  The relation with narrow line luminosity is not
nearly as tight when the extended radio lobe luminosity is used instead of the
total jet power.
We use the optical continuum luminosity as a measure of the available intrinsic
jet power, rather than emission-line luminosity because the latter is known to
have large intrinsic scatter -- determined by differences in covering factor,
distance of ionized gas from the ionizing continuum, and physical conditions.

Thus, we define, as an alternative measure of core dominance, R$_V$, the ratio
of the radio core luminosity at 5GHz rest frequency to the optical V band
luminosity:
log R$_V$ = log(L$_{core}$/L$_{opt}$) = log(L$_{core}$) + M$_{abs}/2.5$ $-$
13.69, where M$_{abs}$ is the absolute magnitude based on K-corrected V
magnitudes.  We use the M$_{abs}$ values conveniently tabulated by
V\'eron-Cetty
\& V\'eron (1993), but correct them to q$_0 = 0$.  Hereinafter we use H$_0 =
50$
km s$^{-1}$\ Mpc$^{-1}$\ and q$_0 = 0$.
In this {\em Letter}, we justify R$_V$\ as an improved measure of orientation,
show an application, and discuss some implications of this new parameter.

\section{Core Dominance vs. Jet Angle}

Figures 1a and 1b compare the correlations of R and R$_V$ vs. the angle between
the jet and the line of sight, $\phi$, for 33 FR II radio sources.  The jet
angle is calculated using apparent superluminal motions from VLBI and the
Doppler factor found from the ratio of radio core flux density to upper limits
on the inverse Compton scattered X-ray flux density.  For 29 sources we use
the data and calculations given by Ghisellini et al. (1993).  We include four
other sources using more recently available apparent expansion speeds
(Vermeulen \& Cohen 1994) and $Einstein$ X-ray data (Wilkes et al. 1994),
deriving $\phi$\ using the method of Ghisellini et al. (1993).
These correlations support the use of core dominance to measure beaming angle.
For R vs. $\phi$, the Spearman rank correlation
coefficient, r$_s$, is $-$0.45, with a less than 0.0025 (one-tailed)
probability, P$_{1t}$, of arising by chance from uncorrelated variables.
Two-tailed probabilities are given in Table 1.  The use of
R$_V$\ significantly reduces the scatter (r$_s$ = $-$0.69, P$_{1t}$
$<$10$^{-5}$);
it can be seen that
the greatest variation in R$_V$ is explained by the relativistic
beaming model, and intrinsic variation in jet luminosity or even $\Gamma$\
must be small by comparison.
While the sample is incomplete, and core flux density is used directly in
calculating R and R$_V$, and indirectly in calculating $\phi$ (through the
limits
on Inverse Compton scattering), the correlation is not strongly affected by
this.
To demonstrate this we tested the corresponding correlations
with the superluminal velocity $\beta_{apparent}$, which is derived
from observational quantities that are
independent of core-dominance.  For R the correlation is significant at the
P$_{1t} \sim 10$\% level, compared with $< 1$\% for R$_V$.  The fact that
R$_V$\ improves the relations for both $\beta_{apparent}$\ and $\phi$, while
both R and R$_V$\ are uncorrelated with $\Gamma$ (Table 1) supports both the
relativistic beaming model and the use of R$_V$\ as an orientation indicator.

\section{Core Dominance vs Broad Line Widths}

Figure 2 shows correlations between the width of the broad H$\beta$
emission line and the measures of core dominance (R and R$_V$)
for two samples of low z quasars investigated by Wills \& Browne (1986,
hereafter WB86) and Brotherton (1995; hereafter B95).
There are $\sim$40 objects in common.
The correlation is significantly improved for both samples when R$_V$ is used
(Table 1).
The improvement with R$_V$ is not the result of a correlation
between line width and optical luminosity; there is no significant correlation
of FWHM with either optical or extended radio luminosity.  We interpret the
improvement with R$_V$ instead of R as the result of avoiding the large
intrinsic scatter of normalizing the core luminosity by the extended radio
luminosity.

Because the relationship between FWHM and core-dominance has been significantly
tightened, FWHM must depend on whatever determines the wide range in R$_V$.
The
results of \S3 suggest that this is orientation rather than intrinsic core
power.

Wills \& Browne (1986) attributed the inverse correlation between FWHM and R
to geometric projection of an axisymmetric velocity
field.  Osterbrock (1977) had earlier suggested projection to account for the
wide range in line widths of Seyfert 1 galaxies, despite their similar physical
conditions (line ratios).  More recently Brotherton et al. (1994) found
strong relationships between broad-line ratios and line widths for
QSOs showing that, for UV lines, FWHM is often determined by varying relative
contributions of two or more kinematically distinct emission regions with
different physical conditions.  This was reinforced for H$\beta$ by B95 who
found, using principal component analysis, that the
most straightforward interpretation is that a very broad (9000 km s$^{-1}$),
component is strong in lobe-dominant sources and weak in
core-dominant sources.

\section{Discussion}

The use of optical continuum luminosity rather than extended radio luminosity
to represent the unbeamed jet power probably works better because the latter
is affected by source-to-source differences in jet interaction with an
inhomogeneous intergalactic medium.
Optical continuum luminosity may plausibly be proportional to the intrinsic
 (unbeamed)
power of the radio jets if it arises from accretion onto a massive compact
object whose rotation powers the radio jets.  This hypothesis is consistent
with the improved correlations -- a result that must be included in devising
a model for quasars' central engines.
A corollary to this is that the ratio of
extended radio luminosity to optical continuum luminosity could be a useful
probe of the interstellar medium, the intergalactic medium, or galaxy
cluster environment of AGNs, beyond several Kpc from the central engine.

There are some additional reasons to believe that the new parameter is
an improvement on R.  R$_V$ may be less affected by long-timescale variability,
since radio core and optical luminosity arise within the nucleus, whereas the
extended radio emission is a longterm time-averaged representation of
the central engine power.  Also, R$_V$\ is more easily obtained in many cases.
It is easier to determine an optical luminosity than an extended radio
luminosity for low redshifts, especially where extended lobe emission may be
resolved.  At higher redshifts, a lobe structure is more difficult to
define because of increasing interaction with the environment
(e.g., Barthel, Tytler, \& Thompson 1990), but core luminosity may still be
obtained.  The correlations might be improved by obtaining more accurate
optical photometry, preferably simultaneous with the radio core flux densities.

Figure 3 directly compares R and R$_V$\ using objects from both B95
and WB86.  The two measures differ approximately by a constant factor: log R =
log R$_V^{1.10 \pm 0.08} -$ 2.7 $\pm$ 0.2 (using the ordinary least squares
bisector as computed by Isobe et al. 1989).

There are some limitations to the use of the optical continuum as a
normalization factor.  One should exclude the beamed synchrotron emission
present in most core-dominant quasars.  This correction is not easy to make,
but, fortunately, if broad emission lines are clearly present, as for the
objects
in the samples used here, then the unbeamed continuum dominates, and the
correction is small.
This correction may move some of the highest R points to higher R$_{V}$\
in Figure 3.  The non-synchrotron continuum may not be isotropic, for example,
if it is emitted from from an accretion disk or its corona.  Figure 3 shows R
$\propto$ R$_{V}$, so there is no strongly axisymmetric emission.
One could correct the optical continuum for reddening in cases of dusty
quasars or radio galaxies.  In practice, this may be difficult.  Choosing an
infrared continuum may reduce this problem (e.g., Spignolio \& Malkan 1989).
Also, a monochromatic optical luminosity may not accurately reflect intrinsic
luminosity if there are object-to-object differences in spectral shape
(see Netzer et al. 1995; Spignolio \& Malkan 1989).
In cases where the true optical continuum cannot be determined, such as for
dusty radio galaxies or extreme blazars, one can use R
together with the relation of Fig. 3 to calculate an effective R$_V$.

\acknowledgments
We thank Alan Bridle, John Conway, David Hough, Steve Rawlings, Derek
Wills, and Chris Impey for helpful input.  We are grateful to STScI for
grant GO-2578.
\clearpage
\begin{planotable}{lcc}
\tablewidth{20pc}
\tablenum{1}
\tablecaption{Spearman Rank Corrlelations\tablenotemark{a}}
\tablehead{
\colhead{} & \colhead{R} & \colhead{R$_V$}}
\startdata
$\phi$\ (VLBI sample) & $-$0.45 & $-0.69$ \nl
 & 0.008 & $10^{-5}$ \nl
$\beta_{\rm apparent}$\ (VLBI sample) & 0.32 & 0.46 \nl
 & 0.07 & 0.007 \nl
$\Gamma$\ (VLBI sample) & 0.33 & 0.19 \nl
 & 0.06 & \nodata \nl
FWHM H$\beta$ (B95) & $-$0.37 & $-$0.59 \nl
 & 0.004 & $<10^{-5}$ \nl
FWHM H$\beta$ (WB86) & $-$0.34 & $-$0.53 \nl
 & 0.004 & $<10^{-5}$ \nl
L$_{core}$ (B95) & 0.62 & 0.81 \nl
 & $<10^{-6}$ & $<10^{-6}$ \nl
L$_{core}$ (WB86) & 0.72 & 0.83 \nl
 & $<10^{-6}$ & $<10^{-6}$ \nl
L$_{ext}$ (B95) & $-$0.46 & 0.01 \nl
 & $<10^{-3}$ & \nodata \nl
L$_{ext}$ (WB86) & $-$0.23 & 0.13 \nl
 & 0.05 & \nodata \nl
M$_{abs}$ (B95) & $-$0.04 & $-$0.07 \nl
 & \nodata & \nodata \nl
M$_{abs}$ (WB86) & $-$0.14 & $-$0.08 \nl
 & \nodata & \nodata \nl

\tablenotetext{a}{Spearman rank correlation coefficients are tabulated with
the two-tailed probability of such a correlation arising by chance given
beneath (when $<$ 0.1).  The VLBI sample has 33 objects, B95 has 58, and
WB86, 71 objects.}
\end{planotable}

\clearpage
\centerline{\bf Figure Captions}

\noindent
FIG. 1.\
The core dominance measures R and R$_V$ vs. the beaming angle $\phi$ for
FR II radio sources, from Ghisellini et al. (1993).  Also included are more
recent observations (Vermeulen \& Cohen 1994), for which we have derived
$\phi$\ in the same manner as Ghisellini et al.

\noindent
FIG. 2.\ A comparison of the correlations between Log FWHM H$\beta$
(km s$^{-1}$) and core dominance for both R and R$_V$, for the data
of Brotherton (1995) and the radio-loud quasars of Wills \& Browne (1986)
(a small number of line widths are for Mg II $\lambda$2798 instead of
H$\beta$).
Arrows indicate limits, filled circles indicate log R $>$ 0, and open
circles indicate log R $<$ 0.

\noindent
FIG. 3.\ A direct comparison of R and R$_V$ for the quasars in Fig. 2.
The dotted lines indicate the ordinary least squares (OLS) fits of y on x and
x on y, computed assuming that the limits are detections.

\newpage
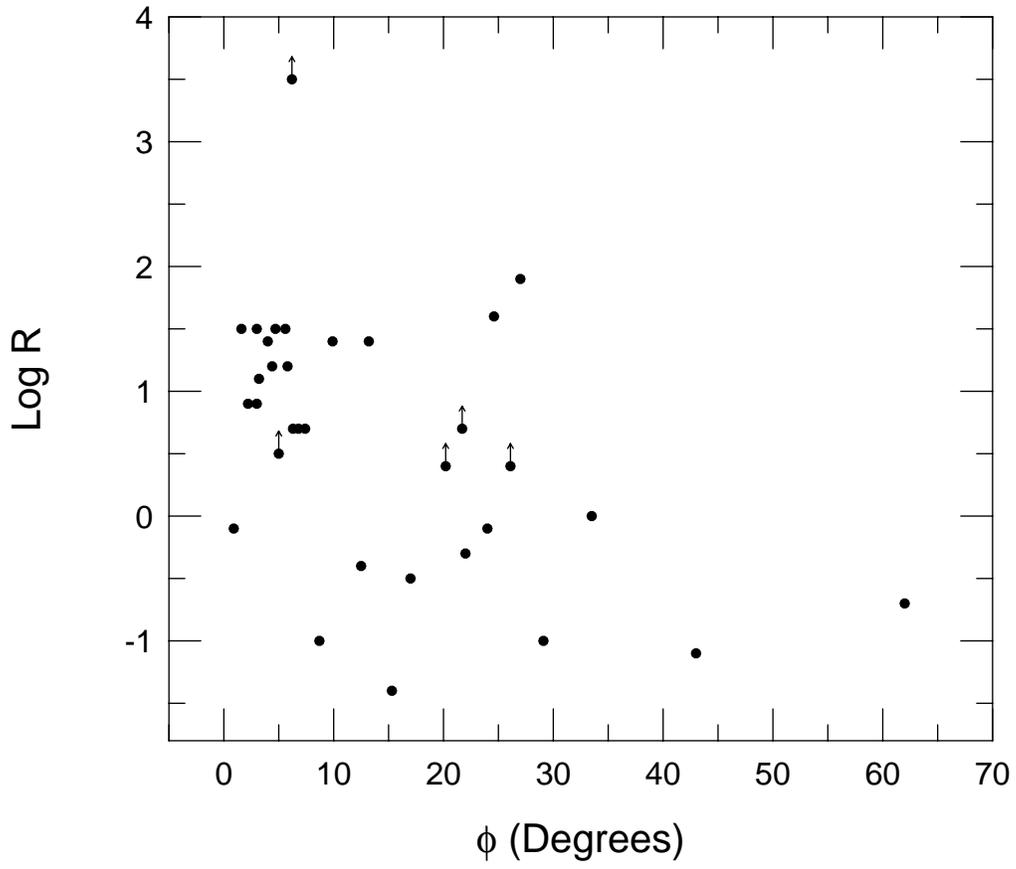
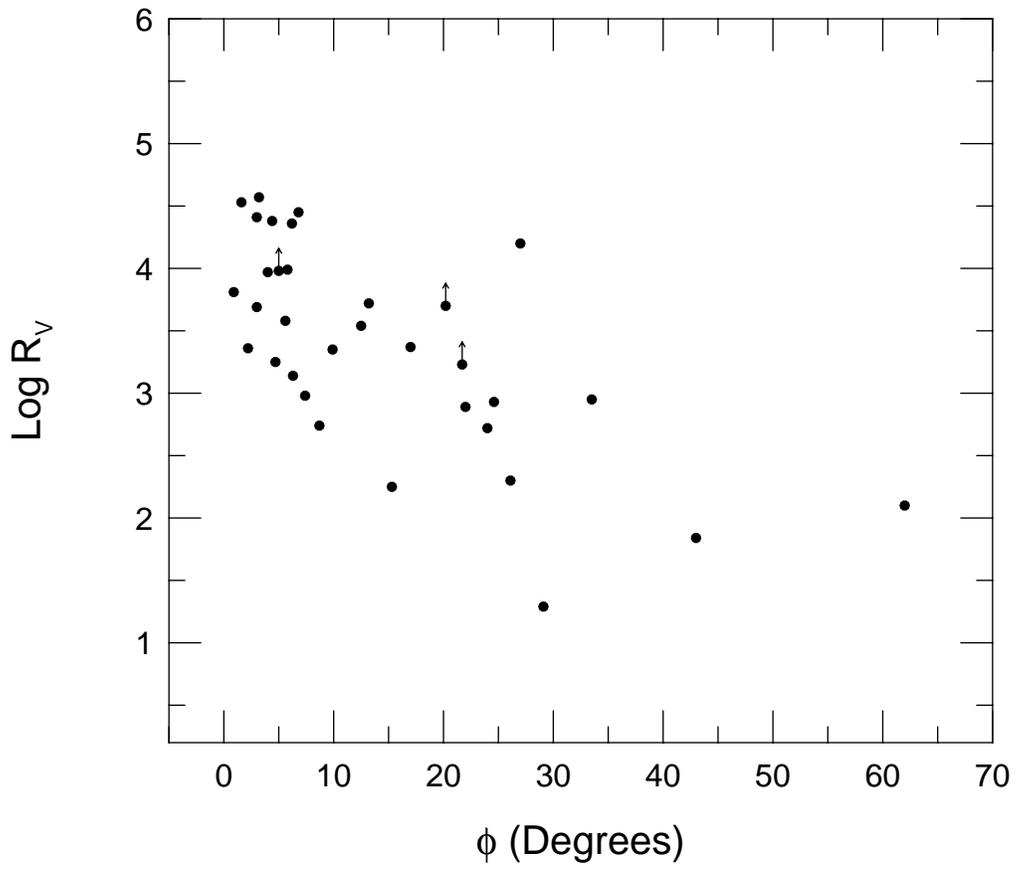
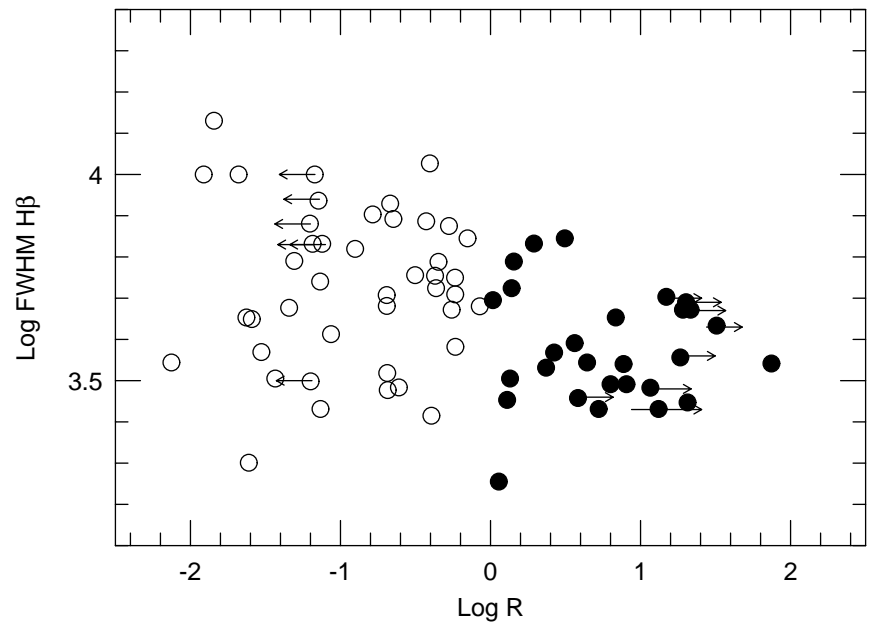
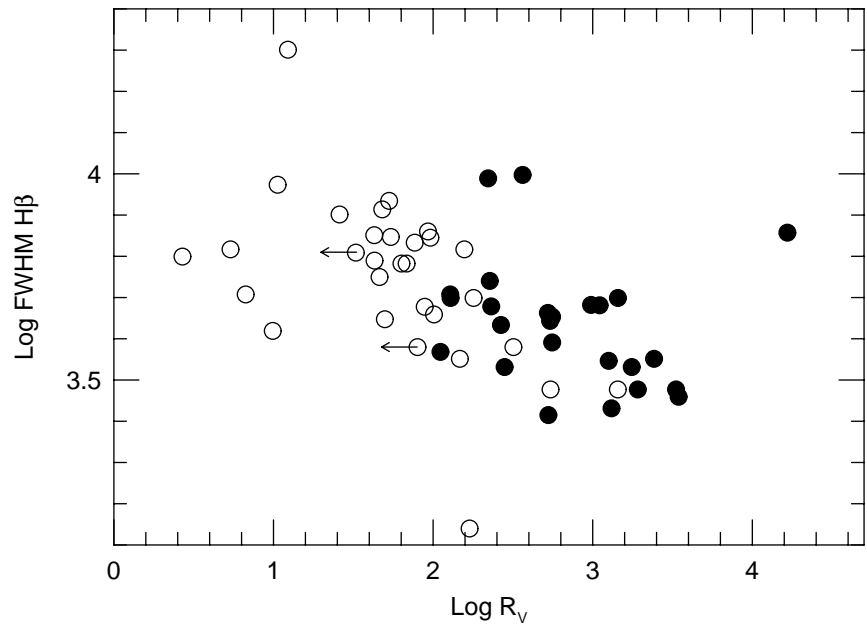
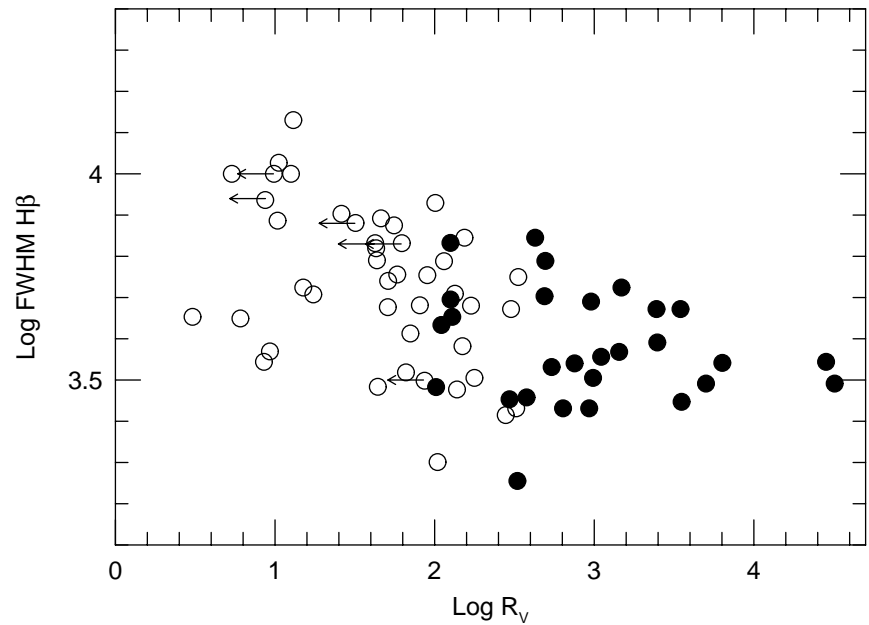
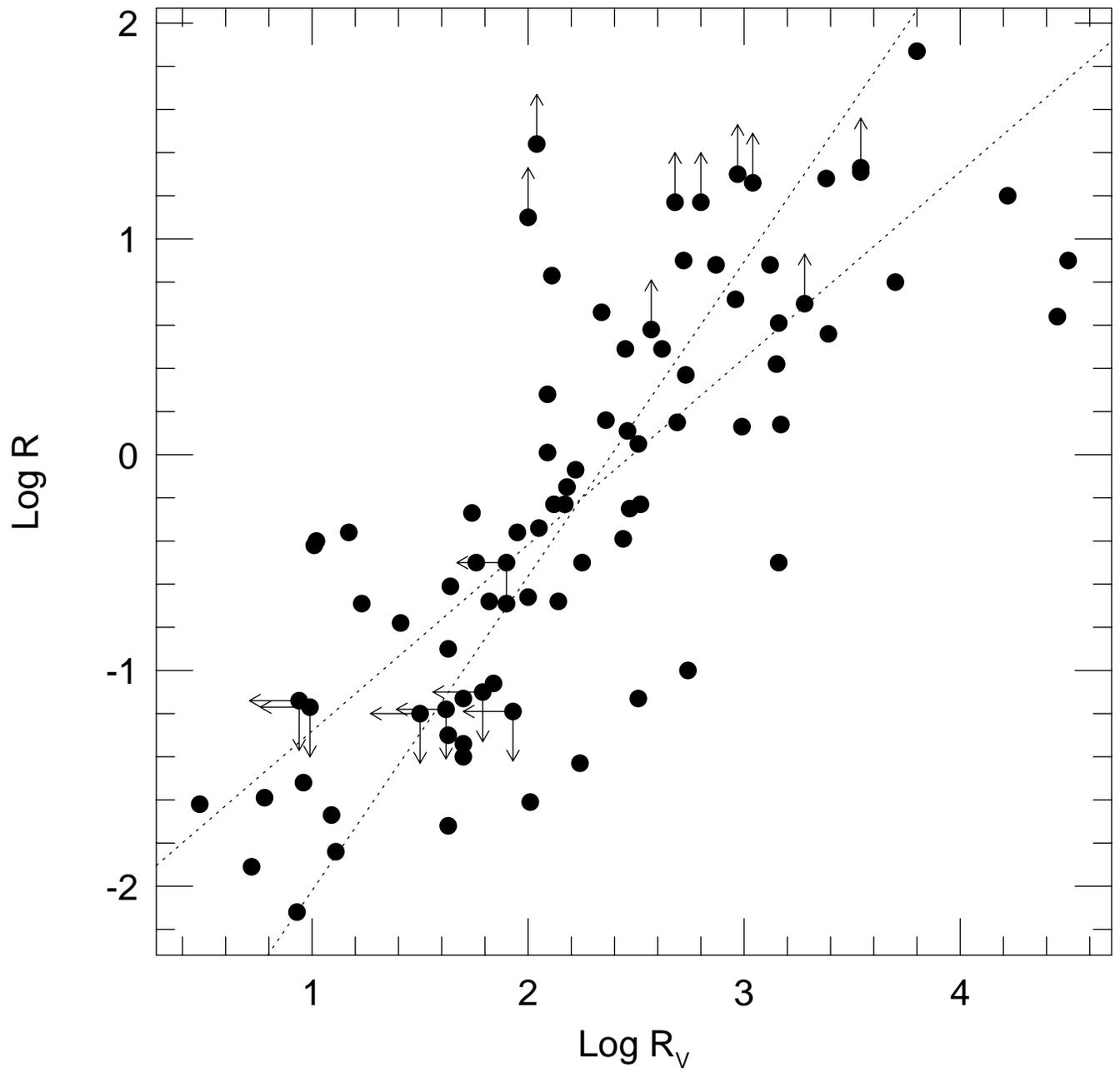
\end{document}